\documentstyle[aps,prl,multicol,amssymb]{revtex}
\title{The power of reduced quantum states}

\author{ N.~Linden$^1$,
S. Popescu$^{2}$, W. K. Wootters$^3$}

\address{
$^1$School of Mathematics, University of Bristol, University
Walk, Bristol BS8 1TW, UK\\
$^2$H.H. Wills Physics Laboratory,
University of Bristol, Tyndall Avenue, Bristol BS8 1TL, UK\\
$^3$Department of Physics, Williams College, Williamstown MA
01267, USA}

\date{18 July 2002}
\begin{document}
\maketitle
\begin{abstract}
In a system of $n$ quantum  particles, we define a measure of the
degree of irreducible $n$-way correlation, by which we mean the
correlation that cannot be accounted for by looking at the states
of $n-1$  particles.  In the case of almost all pure states of
three qubits we show that there is no such correlation:  almost
every pure state of three qubits is completely determined by its
two-particle reduced density matrices.
\end{abstract}


\newcommand{\mtx}[2]{\left(\begin{array}{#1}#2\end{array}\right)}
\newcommand{\tr}{\mbox{Tr} }
\newcommand{\ket}[1]{\left | #1 \right \rangle}
\newcommand{\bra}[1]{\left \langle #1 \right |}
\newcommand{\amp}[2]{\left \langle #1 \left | #2 \right. \right \rangle}
\newcommand{\proj}[1]{\ket{#1} \! \bra{#1}}
\newcommand{\ave}[1]{\left \langle #1 \right \rangle}
\newcommand{\superop}{{\cal E}}
\newcommand{\unity}{\mbox{\bf I}}
\newcommand{\hilbert}{{\cal H}}
\newcommand{\relent}[2]{S \left ( #1 || #2 \right )}
\newcommand{\banner}[1]{\bigskip \noindent {\bf #1} \medskip}

\begin{multicols}{2}
A fundamental question in quantum information theory is to
understand the different types of correlations that quantum states
can exhibit.  A particular issue for a quantum state shared
among $n$ parties, is the extent to which the correlations
between these parties is not attributable to correlations between
groups of fewer than $n$ parties.  In this letter we introduce a
way of characterizing this irreducible $n$-party correlation for
general states of $n$ parties.  Our characterization is based on
measuring the {\em information} in the given quantum
state of $n$ parties that is not already contained in the set of
reduced states of $n-1$ parties.

These considerations lead us to consider the specific case of pure
states of three qubits.  We find the striking result that for
almost all such states, there is no more information in the full
quantum state than is already contained in the three two-party
reduced states.  Expressed differently,  the two-party
correlations uniquely determine the three-party correlations.

In order to explain our construction, let us first treat the case
of states of two parties; the local Hilbert spaces may have any
dimension.  Let the (generally mixed) state be $\rho_{AB}$.  We
ask how much more information there is in $\rho_{AB}$ than is
already contained in the two reduced states $\rho_{A}$ and
$\rho_{B}$.   We address this question by finding another state
$\tilde\rho_{AB}$ which is the most mixed state, {\em i.e.} the
state of maximum entropy, consistent with
the reduced states.  Thus $\tilde\rho_{AB}$ contains all the
information in $\rho_{A}$ and $\rho_{B}$ but no more \cite{Jaynes}.  A simple
calculation using Lagrange multipliers shows that
$\tilde\rho_{AB}$ has the form
\begin{eqnarray}
\tilde\rho_{AB} = \exp(\Lambda_A \otimes 1_B + 1_A\otimes
\Lambda_B).
\label{lambda}
\end{eqnarray}
$1_A$ and $1_B$ denote the identity operators on the Hilbert
spaces of particle $A$ and $B$ respectively. $\Lambda_A$ and
$\Lambda_B$ come from the Lagrange multipliers and  are to be
determined by the condition that the reduced states of
$\tilde\rho_{AB}$ are required to be $\rho_{A}$ and $\rho_{B}$. We
can now solve for the Lagrange multipliers and find that
\begin{eqnarray}
\tilde\rho_{AB} = \rho_{A}\otimes\rho_{B}.
\label{rho}
\end{eqnarray}
In the case that $\rho_{A}$ and $\rho_{B}$ do not have full rank,
this calculation is a little delicate since then the Lagrange
multipliers as they appear in Eq. (\ref{lambda}) are formally infinite.
In that case
we can restrict the Lagrange multipliers to the ranges of $\rho_A$ and
$\rho_B$.  Then Eq.~(\ref{lambda}) defines $\tilde\rho_{AB}$ only on the
subspace in which it is nonzero, but Eq.~(\ref{rho}) remains valid.

The difference $S(\tilde\rho_{AB})-S(\rho_{AB})$, where $S$ is the von
Neumann entropy, can be interpreted as the amount of information
in $\rho_{AB}$ that is not contained in $\rho_A$ and $\rho_B$.  In
fact $S(\tilde\rho_{AB})-S(\rho_{AB})$ is equal to the quantum mutual
information $S(\rho_A)+
S(\rho_B)-S(\rho_{AB})$, which
measures the degree of correlation in $\rho_{AB}$.
(Alternatively, we could use any measure
of the distance between $\tilde\rho_{AB}$ and $\rho_{AB}$ to
express the degree of correlation \cite{Manko}.)
We use the word \lq\lq correlation\rq\rq\ rather
than entanglement since for mixed
states,
$\tilde\rho_{AB}$ will have greater entropy than $\rho_{AB}$  if
$\rho_{AB}$ is separable but not of product form.  For pure states
however $S(\tilde\rho_{AB})=S(\rho_{AB})$ if and only if $\rho_{AB}$ is
of product form, and in this case the difference
$S(\tilde\rho_{AB})-S(\rho_{AB})$
is, except for a factor of two,
the standard measure of bipartite entanglement \cite{BBPS}.
We also note that for a pure state
with reduced states $\rho_{A}$ and $\rho_{B}$,
there are typically many states of two parties having
the same reduced states.  This is in
contrast to the case for more parties, as we will see below.

We now turn to the more interesting case of quantum states of more
than two parties; the local Hilbert spaces may again have any
dimension. For ease of exposition we treat the three-party case
explicitly; the extension to more parties follows
straightforwardly.  Consider, then, a general three-party state
$\rho_{ABC}$. We ask how much more information  there is in
$\rho_{ABC}$ than is already contained in the three reduced states
$\rho_{AB}$, $\rho_{BC}$, $\rho_{AC}$.

Before we analyze this situation we point out that there are a
number of new issues in the three-party case that do not arise in
the two-party case. Consider a set of states $\rho_{AB}$,
$\rho_{BC}$, $\rho_{AC}$ which are supposed to be the reduced
states of some (possibly mixed) state of three parties.  These
three must certainly satisfy some consistency conditions: the
reduced state $\rho_A$ can arise from both $\rho_{AB}$ and
$\rho_{AC}$, and this puts constraints on these two reduced
bipartite states.  However a set of states satisfying this
condition (and the analogous ones for each of the other parties)
may still not correspond to a legitimate state of three parties.  For
consider the following set of reduced states which are supposed to
be the reduced states of some state of three qubits: $\rho_{AB}$,
$\rho_{BC}$, $\rho_{AC}$ are all singlets held between the given
pairs, e.g.
\begin{eqnarray}\rho_{AB} = {1\over\sqrt 2}\left(\ket 0 _A\ket 1 _B -
\ket 1 _A\ket 0 _B\right).
\end{eqnarray}
The reduced states of the individual parties are all the maximally
mixed state of a qubit and so are consistent with each other;
however it is easy to convince oneself that these putative reduced
states are not the reduced state of any three-party state of three
qubits.

We now return to the main theme of our discussion.  We are given a
general three-party state $\rho_{ABC}$.  We argue that a measure
of the irreducible three-party correlations in the state
is the entropy difference between
the state itself and the three-party state that has no more
information in it than in the reduced states.  As in the two-party
case we may use Lagrange multipliers to find the state
$\tilde\rho_{ABC}$ which contains only the information in the
reduced states.  If $\rho_{ABC}$ has maximal rank, then
$\tilde\rho_{ABC}$ is of the form
\begin{equation}
\tilde\rho_{ABC} = \exp(\Lambda_{AB} \otimes 1_C \
  + \Lambda_{AC} \otimes
1_B + \Lambda_{BC}\otimes 1_A).\label{three-party-rho-tilde}
\end{equation}
Here  $\Lambda_{AB}$, $\Lambda_{BC}$ and $\Lambda_{AC}$ come from
the Lagrange multipliers and are to be determined by the condition
that the reduced states of $\tilde\rho_{ABC}$ be those of
$\rho_{ABC}$.  Unlike the case of two parties, we have not been
able to calculate these Lagrange multipliers in closed form, in
general. Nonetheless, the form of $\tilde\rho_{ABC}$ is
illuminating.  For consider a completely general state of three
parties.  It can be expanded using a basis of operators composed
of tensor products of operators spanning each individual Hilbert
space.  For example, for three qubits, a general mixed state may
be written as
\begin{eqnarray}
& &\rho_{ABC} ={1\over 8}\Big( 1 \otimes 1\otimes 1 +\alpha_i
\sigma_i \otimes 1\otimes 1 + \beta_i 1 \otimes \sigma_i \otimes 1
\nonumber\\
& &\quad + \gamma_i 1 \otimes 1 \otimes \sigma_i + R_{ij} \sigma_i
\otimes \sigma_j \otimes 1 + S_{ij} \sigma_i \otimes 1\otimes
\sigma_j \nonumber \\
& &\quad + T_{ij} 1\otimes \sigma_i \otimes \sigma_j  + Q_{ijk}
\sigma_i \otimes \sigma_j \otimes \sigma_k \Big),
\end{eqnarray}
since the set of matrices $(1,\sigma_x,\sigma_y,\sigma_z)$ is a
basis for the operators on ${\mathbb C}^2$.  It is not the case
that the tensor $Q$ describes the three-party correlations (for
consider a density matrix which is of the form
$\rho_{A}\otimes\rho_{B}\otimes\rho_{C}$---it has non-zero $Q$).
However the discussion above shows that for generic density
matrices, a state which has all its information contained in its
reduced states, has the property that its logarithm has no term of
the form $q_{ijk}\sigma_i \otimes \sigma_j \otimes \sigma_k$.

In a number of places in the above discussion we have noted that
the case when the states have non-maximal rank may need careful
treatment.  For example one clearly cannot take the logarithm of
such a state to determine whether its information is contained in
its reduced states.  A particularly important class of states of
non-maximal rank is the set of pure states.  As we will  now see,
this set has surprising properties.

Let us consider the particular case of a system of three qubits.
All  pure states of this system are equivalent under local unitary
transformations to states of the following form \cite{CHS}:
\begin{equation}
|\eta\rangle = a|000\rangle + b|001\rangle + c|010\rangle +
d|100\rangle + e\ket {111}. \label{W}
\end{equation}
The labels within each ket refer to qubits $A$, $B$ and $C$ in
that order; in what follows we will continue to identify qubits
only by the ordering of the labels. We now show that almost all of these states
have no irreducible three-party correlation in the sense
developed in this paper.  That is, we show the following:
except when the parameters
$a,b,c,d,e$ have certain special values, the state $|\eta\rangle$ is the {\em
only} state (pure or mixed) consistent with its two-party
reduced states.

Let $\omega$ be a three-qubit density matrix whose
two-particle reduced states are the same as those of
$|\eta\rangle$. We can think of $\omega$ as obtained from a pure
state $|\psi\rangle$ of a larger system, consisting of the three
qubits and an environment $E$: thus $\omega = \,\hbox{Tr}_E
|\psi\rangle\langle\psi |$. To get a constraint on the form of
$|\psi\rangle$, consider the state $\rho_{AB}$ of qubits $A$ and
$B$ as obtained from $|\eta\rangle$:
\begin{equation}
\rho_{AB} = |\phi_0\rangle\langle\phi_0| +
|\phi_1\rangle\langle\phi_1|,  \label{rhoAB}
\end{equation}
where the unnormalized vectors $|\phi_0\rangle$ and
$|\phi_1\rangle$ are
\begin{eqnarray}
|\phi_0\rangle = a|00\rangle + c|01\rangle + d|10\rangle;\quad
|\phi_1\rangle = b|00\rangle + e\ket{11}. \hfill  \label{phis}
\end{eqnarray}
We insist that $|\psi\rangle$ give this same $\rho_{AB}$ when
restricted to the pair $AB$.  Because $\rho_{AB}$ is confined to
the two-dimensional space spanned by $|\phi_0\rangle$ and
$|\phi_1\rangle$, $|\psi\rangle$ must have the form
\begin{equation}
|\psi\rangle = |\phi_0\rangle |E_0\rangle + |\phi_1\rangle
|E_1\rangle   \label{genpsi}
\end{equation}
where $|E_0\rangle$ and $|E_1\rangle$ are vectors in the
state-space of the composite system consisting of qubit $C$ and
the environment $E$. Computing the density matrix of $AB$ from
Eq.~(\ref{genpsi}) and comparing it with Eq.~(\ref{rhoAB}), we see
that $|E_0\rangle$ and $|E_1\rangle$ must be orthonormal. It will
be helpful to expand $|E_0\rangle$ and $|E_1\rangle$ in terms of
states of $C$ and states of $E$:
\begin{equation}
|E_0\rangle = |0\rangle |e_{00}\rangle + |1\rangle |e_{01}\rangle
; \ |E_1\rangle = |0\rangle |e_{10}\rangle + |1\rangle
|e_{11}\rangle.  \label{Es}
\end{equation}
Here the environment states $|e_{ij}\rangle$ are {\em a priori}
not necessarily either normalized or orthogonal. Combining
Eqs.~(\ref{phis}), (\ref{genpsi}) and (\ref{Es}), we can write
\begin{eqnarray}
& &|\psi\rangle =\big(a|00\rangle + c|01\rangle + d|10\rangle\big)
\big(|0\rangle |e_{00}\rangle + |1\rangle |e_{01}\rangle\big) \nonumber\\
& &\quad +\big(b|00\rangle+e\ket{11}\big) \big(|0\rangle
|e_{10}\rangle + |1\rangle |e_{11}\rangle\big). \label{psi}
\end{eqnarray}
In order to see
what further constraints are imposed on $|\psi\rangle$ by the
requirement that the reduced states agree with $|\eta\rangle$ for
the other pairs, let us consider three specific elements of the
two-party
density matrices.

$\langle 11|\rho_{BC} |11\rangle$: As computed from the state
$|\eta\rangle$, this matrix element has the value $|e|^2$.  As
computed from Eq.~(\ref{psi}), it has the value $|c|^2\langle
e_{01}|e_{01}\rangle + |e|^2\langle e_{11}|e_{11}\rangle$.

$\langle 11|\rho_{AC} |11\rangle$: As computed from
$|\eta\rangle$, this matrix element has the value $|e|^2$. As
computed from Eq.~(\ref{psi}), it has the value $|e|^2\langle
e_{11}|e_{11}\rangle + |d|^2\langle e_{01}|e_{01}\rangle $. Hence
for generic values of $c,d$ and $e$, $|e_{01}\rangle =0$ and
$\langle e_{11}|e_{11}\rangle = 1$, from which it follows that
$|e_{10}\rangle = 0$ and $\langle e_{00}|e_{00}\rangle =
1$.

$\langle 01|\rho_{BC}|10\rangle$:  As computed from
$|\eta\rangle$, this matrix element has the value $bc^*$.  As
computed from Eq.~(\ref{psi}) (with $|e_{01}\rangle =
|e_{10}\rangle = 0$), it has the value $bc^*\langle
e_{00}|e_{11}\rangle$. We conclude, again for generic
values of the parameters, that $|e_{00}\rangle =
|e_{11}\rangle$.

Inserting these inferences into Eq.~(\ref{psi}), we find that
\begin{equation}
|\psi\rangle = \big(a|000\rangle + b|001\rangle + c|010\rangle +
d|100\rangle + e|111\rangle\big)|e_{00}\rangle.
\end{equation}
When we trace out the environment to get the state $\omega$, we
see that we must have $\omega = |\eta\rangle\langle \eta|$.  That
is, the only state ({\em pure or mixed}) consistent with the
two-particle reduced states of $|\eta\rangle$ is $|\eta\rangle$
itself.

The above treatment deals simply with the generic pure state of
three qubits.  We have found it necessary to use a slightly more
involved analysis, to be found in the Appendix, to identify those
special states for which the two-party reduced states do not
uniquely determine the full three-party state. The results in the
Appendix show that the only states that do not have this generic
property are those which are equivalent under local rotations
to states of the form
\begin{equation}
a|000\rangle + b|111\rangle.\label{GHZ}
\end{equation}
We can identify these exceptional states as
the ones
that admit a three-particle Schmidt decomposition as studied
by Peres \cite{Peres}.

The results of this letter clearly raise many questions.  For
example whether the properties that we have found for generic pure
states of three qubits extend to systems of more parties and in
higher dimensional Hilbert spaces; we intend to return to this in
a future publication. Also it is interesting to find non-trivial
classes of $n$-party states that are determined by their reduced
states of {\em fewer} than $n-1$ parties, and to characterize their
entanglement properties. An example is the family of states
\begin{equation}
 a|0001\rangle + b|0010\rangle + c|0100\rangle +
d|1000\rangle.
\end{equation}
These states are uniquely determined by their two-party reduced
states.

Finally, we note that many of the ideas we have put forward here also
shed light on classical probability distributions.  For example
the idea of characterizing the $n$-party correlations using the
information in the $(n-1)$-party marginal distributions.  In the
light of our results on pure states of three qubits it is
intriguing to consider the  case of probability distributions
$P(X,Y,Z)$ of three random variables each of which has two
values; such a distribution arises from local von Neumann measurements on
states of three qubits.  In this case it is not difficult to see
that generic distributions are by no means determined by their
marginal distributions.  For consider a given set of probabilities
$p_{ijk}$ where $p_{000}$ is the probability that $X=0,Y=0,Z=0$
etc.  The set of probabilities $q_{ijk} = p_{ijk} + \delta
(-1)^{\epsilon(ijk)}$ has the same two-party marginal
distributions, where $\delta $ is a constant and $\epsilon(ijk)$
is the parity of the bit string $ijk$.

We thank Dan Collins, Richard Jozsa, Ben Schumacher and Bernard
Silverman for illuminating conversations. We gratefully
acknowledge funding from the European Union under the project
EQUIP (contract IST-1999-11063).

\hrulefill

\leftline{\bf Appendix} Consider an arbitrary pure state
$|\eta\rangle= \sum_{ijk} a_{ijk} |ijk\rangle$ of three qubits $A$,
$B$ and $C$. We give an alternative derivation that for generic
$a_{ijk}$, $|\eta\rangle$ is uniquely determined by its
two-party reduced states and find those states for which this
is not true.

We can quickly dispose of the case in which $|\eta\rangle$ is the
product of a single-qubit state and a two-qubit state. In that
case the two-party reduced states determine both factors in the
product and therefore determine $|\eta\rangle$ uniquely.  In what
follows, we will assume that $|\eta\rangle$ does not have this
product form.

A general state that agrees with $|\eta\rangle$ in its reduced
states can always be obtained from a pure state $|\psi\rangle$
of the three qubits plus an environment $E$.  Let us
first ask what form $|\psi\rangle$ must take in order to be
consistent with the (generally mixed) state of the pair $AB$
derived from $|\eta\rangle$.  By an argument essentially identical
to the one leading to Eq. (\ref{psi}), we find that $|\psi\rangle$
must be of the form
\begin{equation}
|\psi\rangle = \sum_{ijkl} a_{ijl}|ijk\rangle |e_{lk}\rangle.
\label{formAB}
\end{equation}
Here $l$ takes the values 0 and 1, and the states $|e_{lk}\rangle$, which
are states of $E$ alone, satisfy the orthonormality condition
\begin{equation}
\sum_k \langle e_{lk}|e_{l'k}\rangle = \delta_{ll'}.
\label{orthAB}
\end{equation}
Similarly, by considering $AC$ and $BC$ we see that
\begin{equation}
\ket\psi =\sum_{ijkl} a_{ilk}|ijk\rangle |f_{lj}\rangle
=\sum_{ijkl} a_{ljk}|ijk\rangle |g_{li}\rangle, \label{together}
\end{equation}
with $ \sum_j \langle f_{lj}|f_{l'j}\rangle = \delta_{ll'} $ and $
\sum_i \langle g_{li}|g_{l'i}\rangle = \delta_{ll'}. $ Here we
regard the coefficients $a_{ijk}$ as fixed---that is, the state
$|\eta\rangle$ is fixed---and we are looking for environment
vectors $|e_{lk}\rangle$, $|f_{lj}\rangle$ and $|g_{li}\rangle$
that satisfy the various linear equations arising from the fact
that the three expressions for $\ket\psi$ in Eqs. (\ref{formAB})
and (\ref{together}) must all be equal.

It is instructive to write down explicitly, as an example, the two
equations arising from (\ref{formAB}) and (\ref{together}) that
involve only the vectors $|e_{00}\rangle$, $|e_{10}\rangle$,
$|f_{00}\rangle$ and $|f_{10}\rangle$:
\begin{eqnarray}
a_{000}|e_{00}\rangle + a_{001}|e_{10}\rangle =
a_{000}|f_{00}\rangle + a_{010}|f_{10}\rangle
\label{ef0} \\
a_{100}|e_{00}\rangle + a_{101}|e_{10}\rangle =
a_{100}|f_{00}\rangle + a_{110}|f_{10}\rangle\, .
\label{ef1}
\end{eqnarray}
Notice that these two equations are linearly independent: if they
were not, then the state $|\eta\rangle$ would be factorable into
a single-qubit state and a two-qubit state, contrary to our
current assumptions.

These equations and analogous ones relating $|e_{lk}\rangle$ to
$\ket {g_{li}}$ and $\ket {f_{lj}}$ to $\ket {g_{li}}$ can be solved fully, and
one finds that the general solution for the vectors $|e_{lk}\rangle$ is
\begin{eqnarray}
&|e_{01}\rangle = (a_{011}a_{101}-a_{111}a_{001})|v\rangle \,
\nonumber \\
&|e_{10}\rangle = (a_{000}a_{110} - a_{100}a_{010})|v\rangle \,
\nonumber \\
\hfill &|e_{00}\rangle = (a_{100}a_{011} +
a_{101}a_{010})|v\rangle
+ |w\rangle \,  \label{thees} \\
\hfill &|e_{11}\rangle = (a_{000}a_{111} +
a_{001}a_{110})|v\rangle + |w\rangle \, , \nonumber
\end{eqnarray}
the vectors $|v\rangle$ and $|w\rangle$ being arbitrary. The
corresponding expressions for the $f$ and $g$ vectors can be
obtained from Eq.~(\ref{thees}) by permuting indices; for example,
the expression for each $f$ vector is obtained from the expression
for the corresponding $e$ vector by permuting the last two indices
of every $a_{ijk}$ (without changing the vectors $|v\rangle$
and $|w\rangle$).  Thus, once the two vectors $|v\rangle$ and
$|w\rangle$ have been chosen, the solution is determined.  The
form of the solution shows that at most two dimensions of the
environment can ever be used.

We have not yet taken into account the orthonormality conditions
for the environment states.  Let us now consider just
Eq. (\ref{orthAB}) which constrains the $e$ vectors.  It is
helpful to rewrite Eq.~(\ref{thees}) in terms of a new arbitrary
vector $|z\rangle$ that replaces $|w\rangle$:
\begin{eqnarray}
&|e_{01}\rangle = \alpha |v\rangle
&\quad |e_{10}\rangle = \beta |v\rangle  \nonumber \\
&|e_{00}\rangle = |z\rangle &\quad|e_{11}\rangle = |z\rangle +
\gamma |v\rangle \, , \label{newes}
\end{eqnarray}
where $\alpha = a_{011}a_{101}-a_{111}a_{001}$, $ \beta =
a_{000}a_{110} - a_{100}a_{010}$ and $\gamma = a_{000}a_{111} +
a_{001}a_{110} - a_{100}a_{011} - a_{101}a_{010}$. In  terms of
these parameters, the orthonormality condition of
Eq.~(\ref{orthAB}) is expressed by the following three equations
\begin{eqnarray}
&\langle z|z\rangle + |\alpha|^2 \langle v|v\rangle = 1 \nonumber
\\
&\langle z|z\rangle + (|\beta|^2+|\gamma|^2)\langle v|v\rangle
+\gamma\langle z|v\rangle + \bar{\gamma}\langle v|z\rangle = 1
\label{raworth} \\
&\bar{\alpha}\gamma\langle v|v\rangle + \beta\langle z|v\rangle
+\bar{\alpha}\langle v|z\rangle = 0. \nonumber
\end{eqnarray}
Taking the difference between the first two of these equations,
and treating separately the real and imaginary parts of the third,
we obtain three homogeneous linear equations for the three real
variables $\langle v|v\rangle$, Re\,$\langle z| v\rangle$ and
Im\,$\langle z|v\rangle$.  For generic values of
$\alpha,\ \beta$ and $\gamma$, these three equations are linearly
independent, so that the only solution is $|v\rangle = 0$. This in turn
implies, by Eq.~(\ref{thees}), that $|e_{01}\rangle =
|e_{10}\rangle = 0$ and $|e_{00}\rangle = |e_{11}\rangle$.  Thus
in this generic case only a single dimension of the environment is
used---that is, the environment is in a pure state---and the
qubits $ABC$ {\em must} be in the given state $|\eta\rangle$.

This conclusion can be avoided only if the determinant $D$ of
the $3\times 3$ matrix associated with the three homogeneous linear
equations vanishes, and the corresponding determinants computed
from the two other orthonormality conditions (for the vectors $f$
and $g$) are also zero. Computing $D$ explicitly, we find that
$D=0$ if and only if (i) $|\alpha|=|\beta|$ and (ii)
$\bar{\gamma}^2\alpha\beta$ is real and non-negative.

Suppose now that $|\eta\rangle$ is {\em not} determined by its two-party
reduced states, so that the above conditions (i) and (ii)
must be satisfied.  These conditions imply that there exists a local
rotation on qubit $C$ that will bring both $\alpha$ and $\beta$
to zero, thus bringing $|\eta\rangle$ to the form
\begin{equation}
|p_A\rangle |p_B\rangle |0\rangle + |q_A\rangle
|q_B\rangle |1\rangle.
\label{nearGHZ}
\end{equation}
Here $|p_A\rangle$ and $|q_A\rangle$ are (unnormalized) vectors
in the space of qubit $A$, and $|p_B\rangle$ and $|q_B\rangle$
belong to qubit $B$.
We now use in a similar way the conditions analogous to (i) and (ii)
but derived from the orthonormality relations for the $f$ vectors.
These imply that we can apply to the form (\ref{nearGHZ}) a local
rotation on qubit $B$ to bring it to the form
$|p_A\rangle |0\rangle |0\rangle + |q_A\rangle
|1\rangle |1\rangle$.
Finally, from the conditions derived from the $g$ vectors, it follows that
we can
rotate qubit $A$ and arrive at the form $a|000\rangle + b|111\rangle$.

We conclude, then, that the only pure three-qubit states that
might {\em not} be uniquely determined by their two-particle
reduced states are those that are equivalent under local rotations to the form
given in Eq.~(\ref{GHZ}). In fact it is easy to see that for any
state of this form with $a\neq 0$ and $b \neq 0$, there do exist
other three-qubit states---{\em e.g.} mixtures of $|000\rangle$
and $|111\rangle$---having the same two-party reduced states.

\end{multicols}
\end{document}